\documentclass[10pt,letterpaper]{article}
\usepackage{opex3}
\newcommand{\rcoh}[1]{\vert #1 \rangle\langle #1 \vert}

\begin{document}

\title{Manipulating the non-Gaussianity of phase-randomized coherent states}

\author{Alessia Allevi,$^{1}$ Stefano Olivares,$^2$ and Maria Bondani$^{3}$}

\address{$^1$Dipartimento di Scienza e Alta Tecnologia, Universit\`a degli Studi dell'Insubria\\ and CNISM UdR Como, via Valleggio 11, 22100 Como, Italy;\\
$^2$Dipartimento di Fisica, Universit\`a degli Studi di Milano and CNISM UdR Milano Statale, via Celoria 16, 20133 Milano, Italy;\\
$^3$Istituto di Fotonica e Nanotecnologie, CNR and CNISM UdR Como, via Valleggio 11, 22100 Como, Italy \\}

\email{alessia.allevi@uninsubria.it} 



\begin{abstract}
We experimentally investigate the non-Gaussian features of the phase-randomized coherent states, a class of states exploited in communication channels and in decoy state-based quantum key distribution protocols. In particular, we reconstruct their phase-insensitive Wigner functions and quantify their non-Gaussianity. The measurements are performed in the mesoscopic photon-number domain by means of a direct detection scheme involving linear detectors.
\end{abstract}



\section{Introduction} \label{intro}

Recently, weak phase-randomized or phase-averaged coherent states (PHAVs) have been successfully exploited to implement quantum key distribution (QKD) \cite{PRL:05,APL:07,EPJD}. It has been demonstrated that phase randomization is useful to improve the security of QKD, pointing out the relevance of a careful characterization of the employed sources \cite{CALT:05,curty10}. With respect to coherent states, which are described by Gaussian Wigner functions \cite{oli:12}, PHAVs clearly exhibit non-Gaussian features. Thus, the systematic study of the nature of PHAVs and the possibility to manipulate it can be considered of real interest in enhancing the performances of the communication protocols in which such states are used \cite{curty09}.
The complete characterization of optical states can be obtained by reconstructing the Wigner function.
In practice, two alternative detection techniques, both based on interferometric schemes, can be used to achieve this goal. The typical method employed so far is the time-resolved optical homodyne tomography (see Ref.~\cite{lvovsky09} for a review).  Nevertheless, on the basis of the results derived in Ref.~\cite{cahill69}, it has been demonstrated that also direct detection can be used to obtain the Wigner function starting from photon-number distributions \cite{wallentowitz96,banaszek96}.
\par
In this paper we report on the controlled generation of PHAVs \cite{zambra07} and on their full characterization in terms of Wigner functions by exploiting a direct detection scheme involving linear detectors. In particular, we present a systematic study of their non-Gaussianity (nonG) in different experimental conditions and for different choices of the parameters that determine the states. Moreover, we demonstrate that it is possible to manipulate the nonG by linear passive devices (beam splitters). In order to quantify the nonG of the states we adopt a recently proposed measure given in terms of the von Neumann entropy ~\cite{GenoniPRA,genoni08}. This investigation is performed in the mesoscopic photon-number regime as in this domain the optical states are robust with respect to losses and thus it is possible to better understand the influence of the different parameters on the nature of PHAVs.
\par
The paper is structured as follows. In Section~\ref{s:theory} we introduce the concepts and formalism used throughout the paper. In particular we write the theoretical Wigner functions of PHAVs and of the state obtained by mixing two PHAVs at a beam splitter. Section~\ref{s:result} describes our experimental setup for the generation, manipulation and characterization of PHAVs and contains the experimental reconstruction of the Wigner functions of the generated states and the results we obtained for the nonG. We close the paper by drawing  some concluding remarks in Section~\ref{s:remarks}.

\section{Theory}\label{s:theory}
The density matrix $\varrho_\beta$ of a single-mode PHAV of amplitude $\beta$ is given by:
\begin{equation} \label{PHAV}
\varrho_{\beta} = \int_{0}^{2 \pi} \frac{d \phi}{2 \pi} \rcoh{\beta} = e^{-|\beta|^2}\, \sum_{n=0}^{\infty} \frac{|\beta|^{2n}}{n!} \rcoh{n},
\end{equation}
where $|\beta\rangle$ is a coherent state, $\beta = |\beta|\,e^{i \phi}$, and $\{|n\rangle\}$, $n=0,1,2,\ldots$, is the photon number basis. The corresponding Wigner function is \cite{OLwigner}:
\begin{equation}
W_{\rm PHAV}(\alpha;\beta) = \int_{0}^{2 \pi} \frac{d \phi}{2 \pi}\, e^{-|\alpha - \beta|^2}= \frac{2}{\pi} \, I_0(4|\alpha||\beta|)\, \exp[-2(|\alpha|^2+|\beta|^2)]\,,
\label{eq:wignerPHAV}
\end{equation}
$I_0(z)$ being the modified Bessel function. The representation in terms of Wigner function emphasizes both the non-Gaussian nature of PHAV and its phase-insensitive nature ({\em i.e.} rotational invariance about the origin of the phase space).
\par
In order to assess the nonG of the state $\varrho_\beta$, we consider the following measure \cite{GenoniPRA}:
\begin{equation}\label{nonG}
\delta[\varrho_{\beta}] = S(\sigma) - S(\varrho_{\beta}),
\end{equation}
where $S(\varrho) = -\hbox{Tr} [\varrho \ln \varrho]$ is the von Neumann entropy of the state $\varrho$ and $\sigma$ is a reference Gaussian state with the same covariance matrix as the state $\varrho$.  As the PHAV is a diagonal state in the photon number basis, its reference state is a thermal equilibrium state, with the same mean number of photons $N = |\beta|^2$ \cite{GenoniPRA}. In this case Eq.~(\ref{nonG}) reduces to:
\begin{equation}\label{nonGPHAV}
\delta[\varrho_\beta] = (N+1) \ln (N+1)-N \ln N +\sum_n p_n \ln p_n,
\end{equation}
where $p_n =  \exp{(-|\beta|^2)}\, |\beta|^{2n}/n!$.
\par
Another interesting state we will investigate is that obtained from the interference of two PHAVs $\varrho_{\beta}$ and $\varrho_{\tilde{\beta}}$ at a beam splitter (BS) with transmissivity $\tau$, which can be useful for application to passive decoy state QKD \cite{curty09}. In this case, if we consider only one of the two modes outgoing the beam splitter, we still obtain a state with a diagonal density matrix whose Wigner function can be written as follows~\cite{oli:12}:
\begin{equation} \label{eq:wigner2PHAV:B}
W_{\rm 2-PHAV}(\alpha;\beta,\tilde{\beta},\tau) = \int_{0}^{2 \pi} \frac{d\tilde{\phi}}{2\pi}\,
W_{\rm PHAV}(\alpha-\tilde{\beta}\sqrt{1-\tau};\beta\sqrt{\tau})
\end{equation}
where $\tilde{\beta} = |\tilde{\beta}|\,e^{i \tilde{\phi}}$ and the function in the integral is given by Eq.~(\ref{eq:wignerPHAV}). Obviously, the Wigner function of the other outgoing mode can be obtained by replacing $\tau$ with the reflectivity $(1-\tau)$. In the following we will refer to this state as 2-PHAV and, since it is diagonal, its nonG is still given by Eq.~(\ref{nonGPHAV}), where now $p_n$ is the 2-PHAV non-trivial photon statistics reported in \cite{zambra07}.

\section{Experimental results}\label{s:result}
The measurements we performed were aimed at reconstructing the Wigner functions of PHAVs and 2-PHAVs and at comparing their nonG for different choices of the experimental parameters.
The generation of the class of the PHAVs was achieved by exploiting the second harmonics (@ 523 nm, 5-ps pulses) of a mode-locked Nd:YLF laser
amplified at 500 Hz (High-Q Laser Production).
\begin{figure}[htbp]
\centering\includegraphics[width=0.65\textwidth]{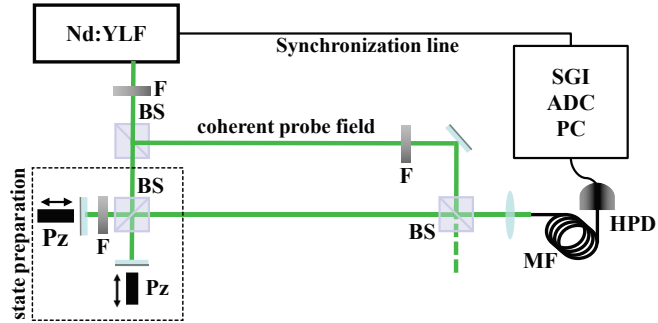}
\caption{Experimental setup. F: variable neutral density filter; BS: 50/50 beam splitter; Pz: piezoelectric movement; MF: multimode fiber (600~$\mu$m core).} \label{setup}
\end{figure}
According to the experimental setup sketched in Fig.~\ref{setup}, the single PHAV was obtained by sending the light pulses to a mirror mounted on a piezo-electric movement. Its displacement, which was controlled by a function generator, was operated at a frequency of 100~Hz and covered a 12~$\mu$m span. The 2-PHAV was produced by mixing at a BS two single PHAVs obtained by duplicating the phase-randomization system. A filter inserted in the path of one of the two PHAVs allowed us to change the balancing between the two fields. We implemented a direct detection scheme involving a photon-counting detector, namely a hybrid photodetector (HPD, R10467U-40, maximum quantum efficiency $\sim$0.5 at 500 nm, 1.4 ns response time, Hamamatsu), that is a detector not only endowed with a partial photon-counting capability, but also characterized by a linear response up to 100 photons. Thanks to its features, the HPD can actually operate in the mesoscopic domain, where the states are robust with respect to losses.  The output of the detector was amplified (preamplifier A250 plus amplifier A275, Amptek), synchronously integrated (SGI, SR250, Stanford) and digitized (AT-MIO-16E-1, National Instruments). The estimated overall quantum efficiency of the detection chain is equal to 0.45. By using the strategy described in Refs.~\cite{JMOself,andreoni09}, we can reconstruct the detected-photon distributions that, as we have already proven in \cite{OLwigner}, represent the basic elements to retrieve the Wigner function.
\par
The experimental reconstruction of this function was achieved by mixing at a BS the state under investigation, in the present case either the single PHAV or the 2-PHAV, with a coherent probe field whose amplitude and phase could be continuously varied. As both the states to be characterized are phase-insensitive, we reconstructed only a section of the Wigner distribution for fixed phase.
\begin{figure}[htbp]
\centering\includegraphics[width=0.45\textwidth]{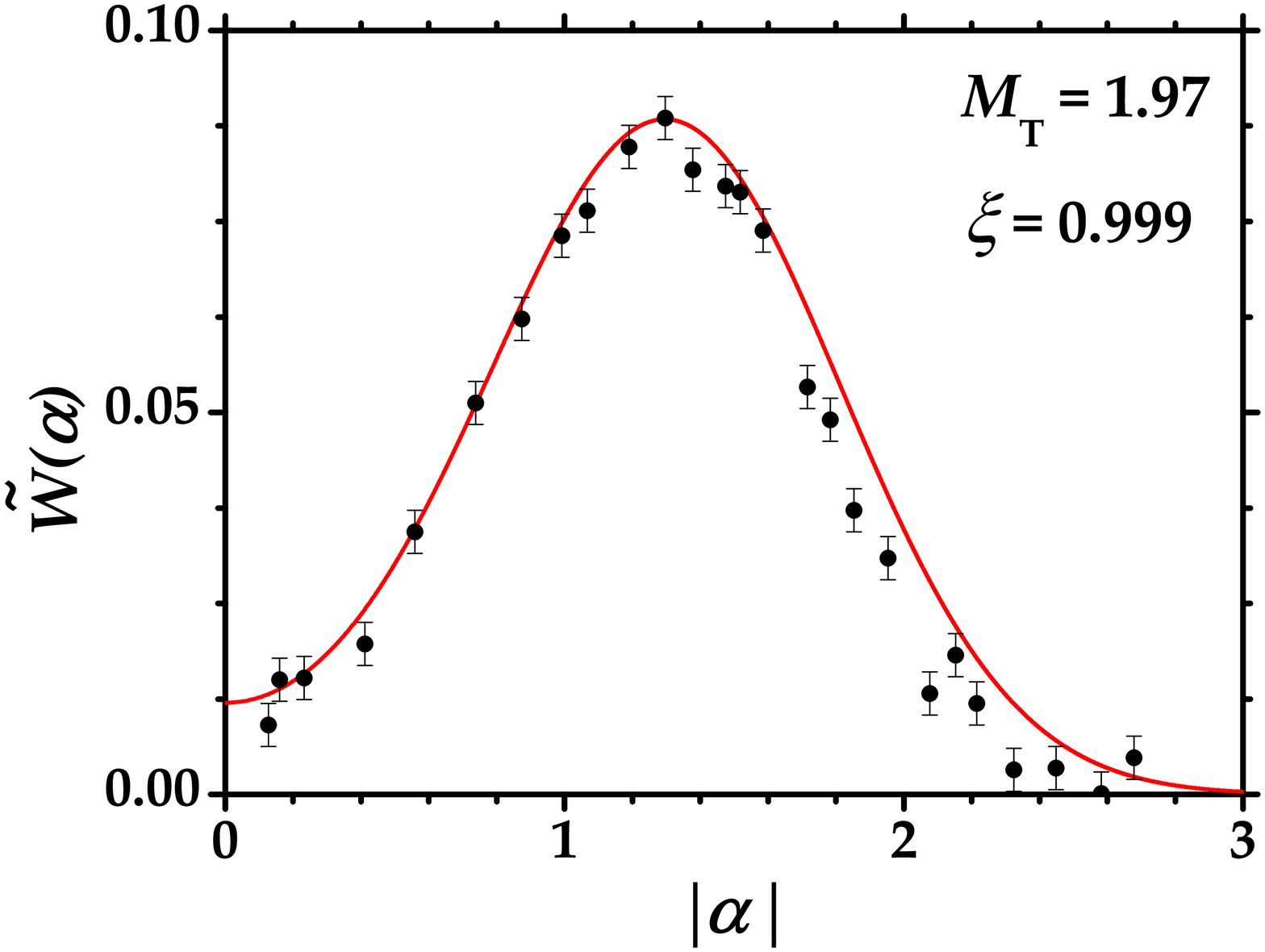} \centering\includegraphics[width=0.45\textwidth]{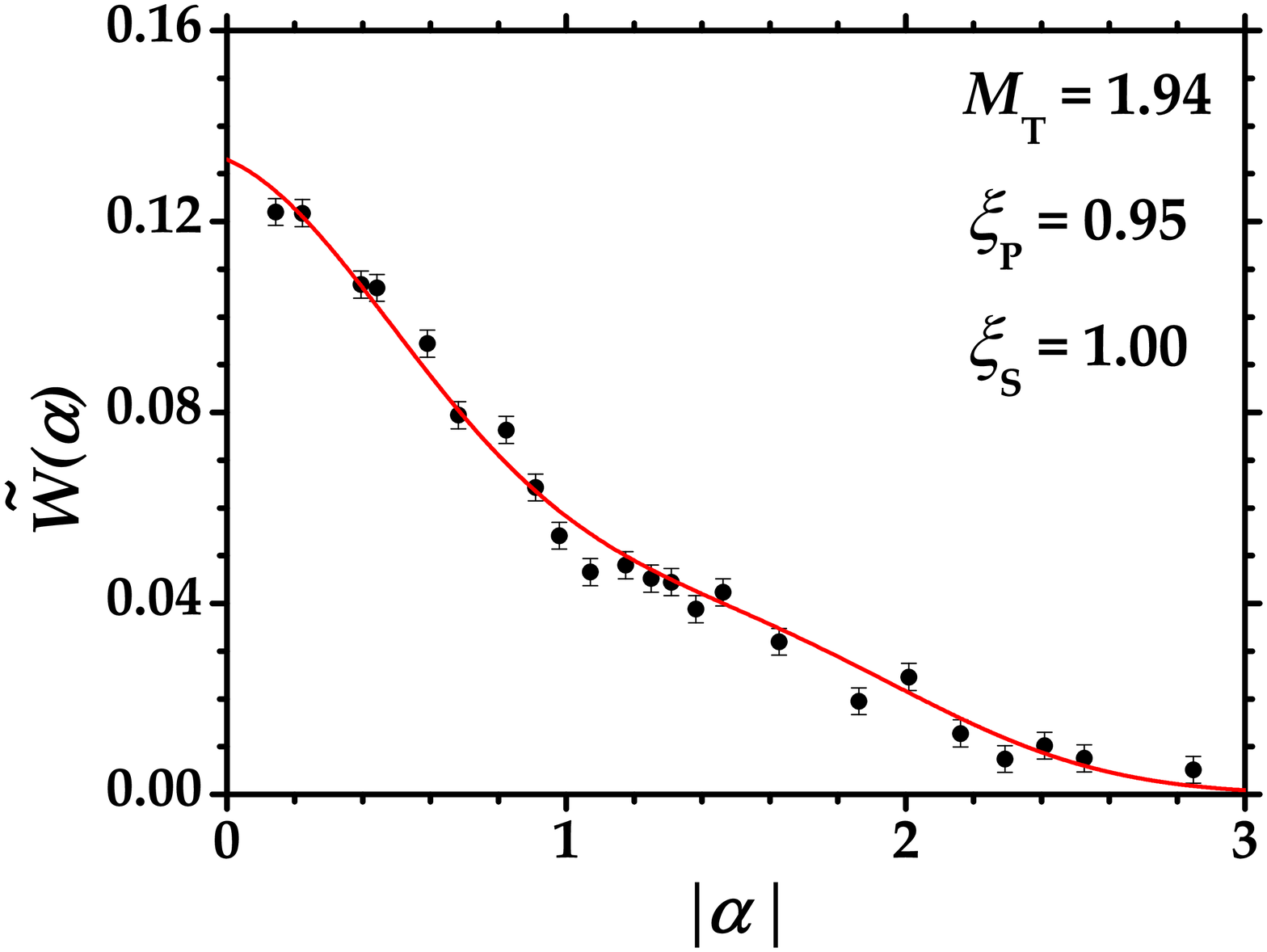}
 \caption{Experimental reconstruction of a section of the phase-insensitive Wigner function of a PHAV (left panel), with $|\beta|^2=1.97$, and of a 2-PHAV (right panel), with $|\beta|^2=1.03$ and $|\tilde{\beta}|^2=0.91$. Black dots: experimental data; red line: theoretical expectations given in Eqs.~(\ref{eq:wignerPHAVoverlap}) and (\ref{eq:wigner2PHAVoverlap}), respectively.} \label{wig-comp}
\end{figure}
In Fig.~\ref{wig-comp} we plot the experimental data (black dots + error bars) of a single PHAV (left panel) and of a balanced 2-PHAV (right panel), endowed with nearly the same mean number of detected photons, $M_{\rm T}=1.97$ and $M_{\rm T}=1.94$, respectively. In each panel we also show the theoretical expectations (red line) for the PHAV and 2-PHAV, respectively \cite{OLwigner}:
\begin{equation}
\tilde{W}_{\rm PHAV}(\sqrt{\xi} \alpha) = W_{\rm PHAV}(\sqrt{\xi} \alpha) e^{-\sqrt{1-\xi}(|\alpha|+|\beta|)}, \label{eq:wignerPHAVoverlap}
\end{equation}
$\xi$ being the overall (spatial and temporal) overlap between the probe and the PHAV, and:
\begin{equation}
\tilde{W}_{\rm 2-PHAV}(\sqrt{\xi_{\rm P}} \alpha) = W_{\rm 2-PHAV}(\sqrt{\xi}_{\rm P} \alpha) e^{-[\sqrt{1-\xi_{\rm P}}|\alpha|+\sqrt{1-\xi_{\rm S}}(|\beta|+|\tilde{\beta}|)]}, \label{eq:wigner2PHAVoverlap}
\end{equation}
where $\xi_{\rm P}$ describes the overall overlap between the probe and the 2-PHAV and $\xi_{\rm S}$ the overall overlap between the two components of the 2-PHAV. In both the Eqs.~(\ref{eq:wignerPHAVoverlap}) and (\ref{eq:wigner2PHAVoverlap}), $|\beta|^2$ and $|\tilde{\beta}|^2$ are now the mean numbers of photons we measured (see Fig.~\ref{wig-comp}), thus including the quantum efficiency. In fact, it is worth noting that for classical states the functional form of the Wigner function is preserved also in the presence of losses and its expression, given in terms of detected photons, reads $ \tilde{W}(\alpha) =\frac{2}{\pi} \sum_{m=0}^\infty (-1)^m p_{m,\alpha}^{\ el}$, where $p_{m,\alpha}^{\ el}$ represent the detected-photon distributions of the state to be measured displaced by the probe field \cite{OLwigner}. As testified by the very high values of the overlaps $\xi$, $\xi_{\rm P}$ and $\xi_{\rm S}$ reported in the panels of Fig.~\ref{wig-comp}, we actually achieved a very good superposition in aligning the system. In our opinion, having overcome this criticality can be considered a demonstration of the robustness of our setup and also a fundamental step on the way to the exploitation of these states for possible applications.
\par
\begin{figure}[htbp]
\centering\includegraphics[width=0.45\textwidth]{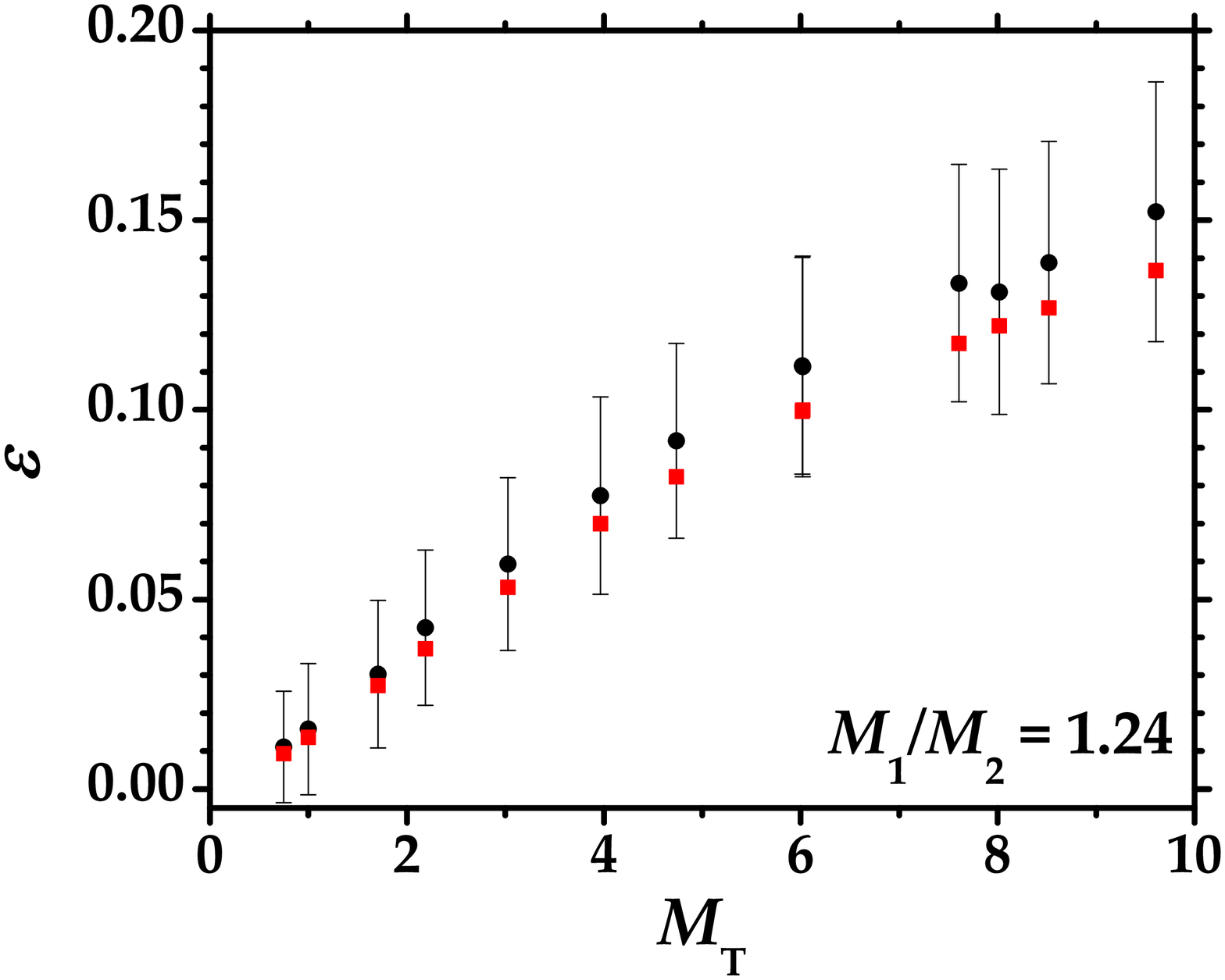} \centering\includegraphics[width=0.45\textwidth]{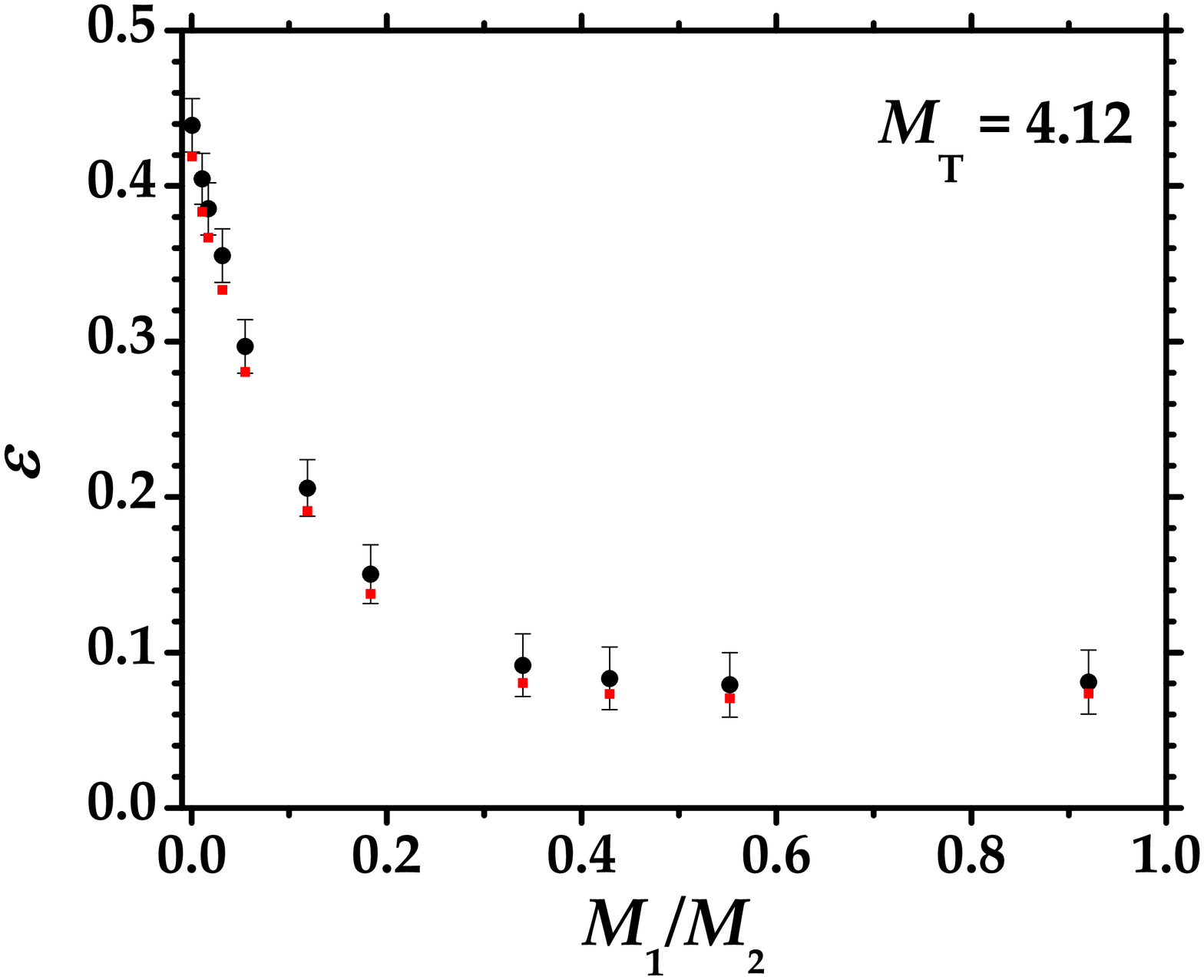}
\caption{Left panel: nonG measure $\epsilon$ for the detected number of photons as a function of the mean number of photons of balanced 2-PHAVs (see the text for details).  Right panel: $\epsilon$ as a function of the balancing between the two components of the 2-PHAV, at fixed mean number of detected photons $M_T = M_1 + M_2$ of the overall state. Black dots: experimental data; red squares: theoretical expectation.} \label{nonGunbal}
\end{figure}

We characterized the nonG amount of the reconstructed states by considering the measure in Eq.~(\ref{nonGPHAV}), which, however, is actually given in terms of photons. As we have not directly access to the number of photons, we calculated a similar expression, $\epsilon$, for detected photons, which represents a lower bound to nonG \cite{GenoniPRA,allevi10}. We found $\epsilon_{\rm PHAV} = 0.207 \pm 0.004$ and $\epsilon_{\rm 2-PHAV} = 0.036 \pm 0.005$. From this comparison it emerges that a PHAV is more non-Gaussian than the 2-PHAV, thus demonstrating that a Wigner function exhibiting a dip in the origin of the phase space is more non-Gaussian than one characterized by a peak in the origin followed by a ``shoulder''. The results prove that combining two non-Gaussian states does not necessarily lead to an increase of the overall nonG. Indeed, to better understand the nature of the ``shoulder", we generated balanced 2-PHAVs at different mean values and reconstructed their Wigner function.
\begin{figure}[htbp]
\centering\includegraphics[width=0.45\textwidth]{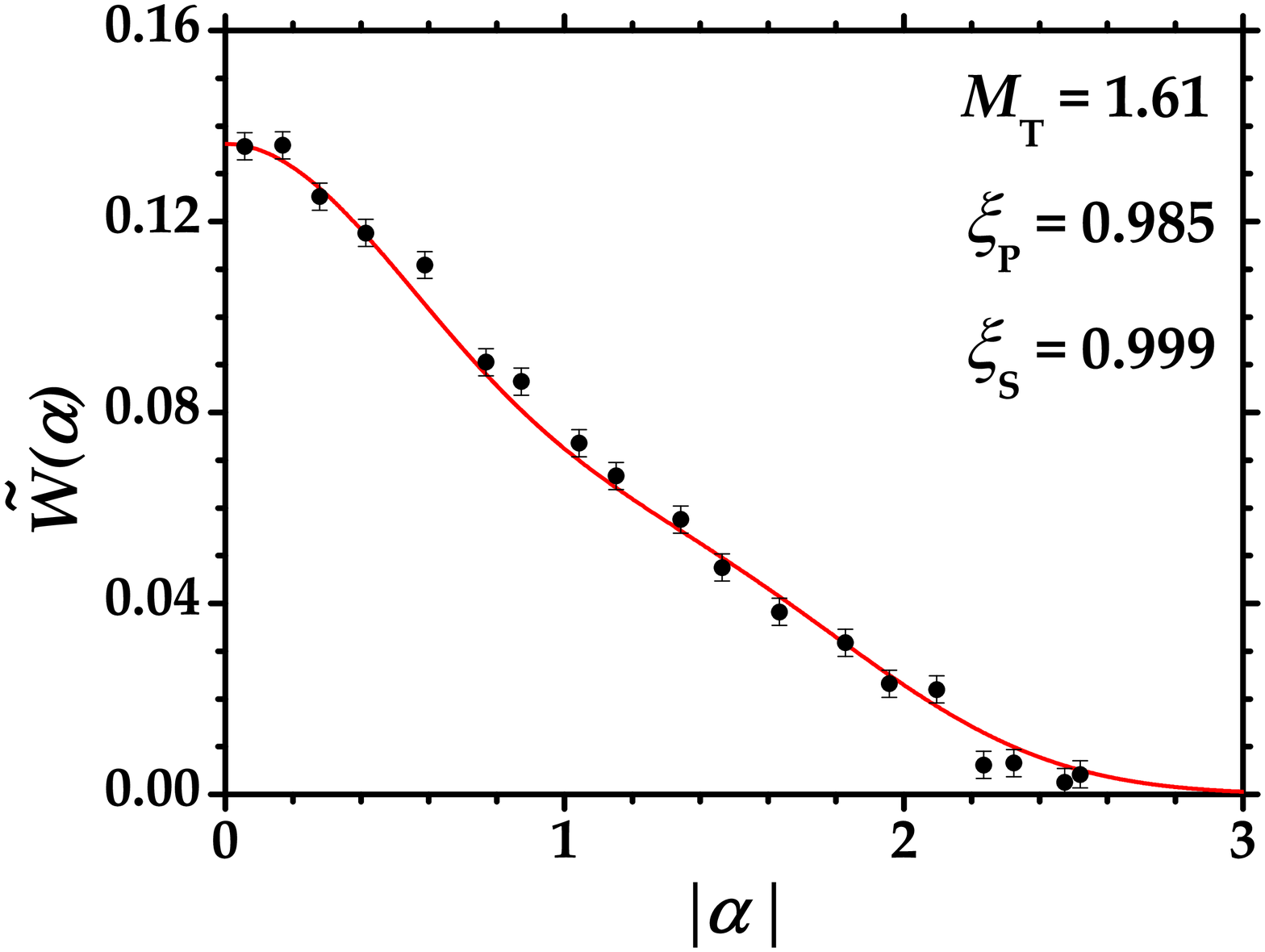}
\centering\includegraphics[width=0.45\textwidth]{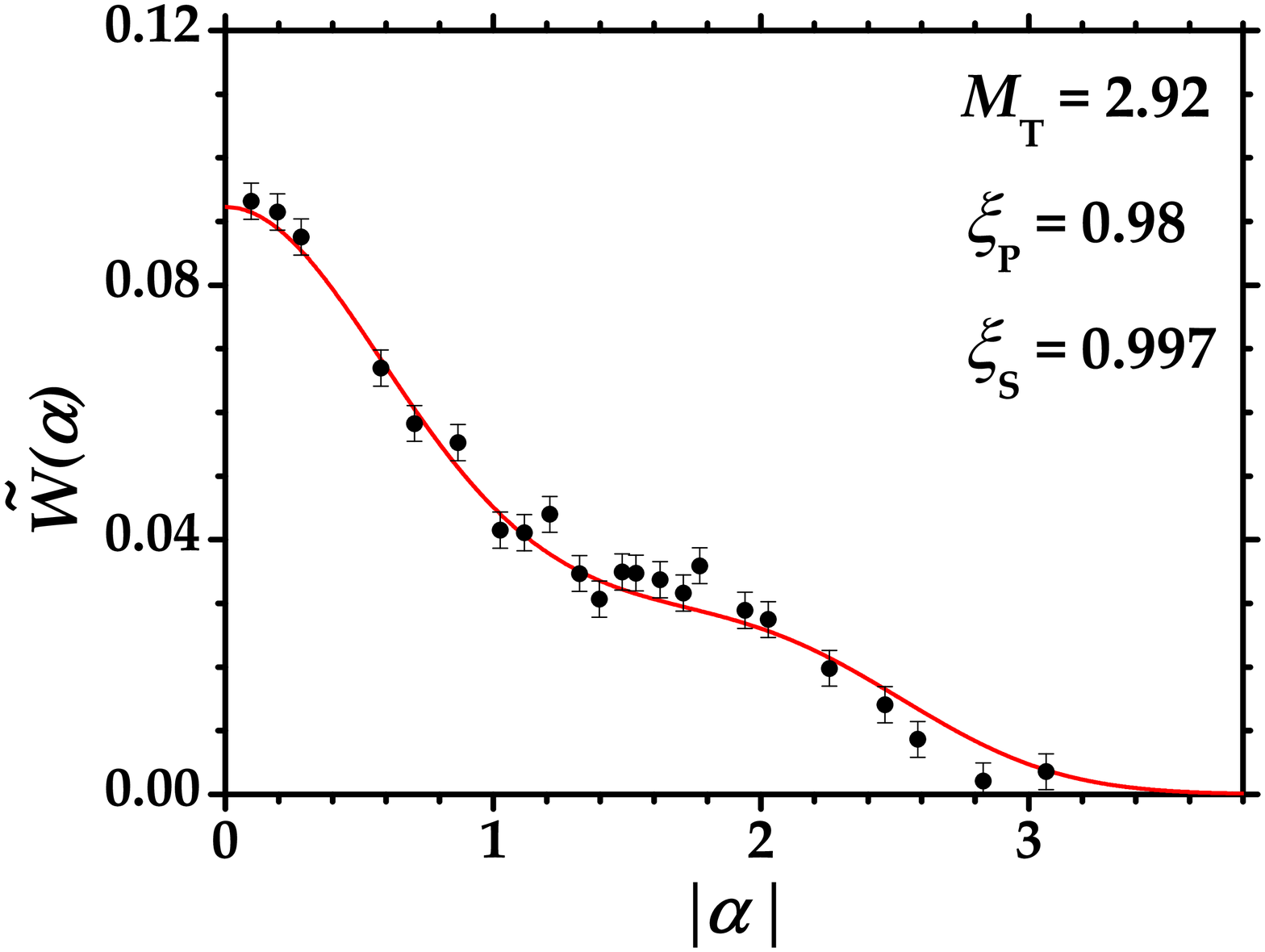}
 \caption{Experimental reconstruction of a section of the phase-insensitive Wigner function of a 2-PHAV at different mean values $M_{\rm T}$ (see the text for details). Black dots: experimental data; red line: theoretical expectations given in Eq.~(\ref{eq:wigner2PHAVoverlap}).} \label{wig-meanvalues}
\end{figure}
The experimental data are shown in Fig.~\ref{wig-meanvalues} together with the theoretical curve. We can observe that the ``shoulder" is more and more evident at increasing the mean number of photons of the balanced 2-PHAV.  Indeed, the calculation of the nonG amount is in accordance with this result, as it is an increasing function of the mean number of photons (left panel $\epsilon_{\rm 2-PHAV} = 0.031 \pm 0.005$, $M_{\rm T}=1.61$, and right panel $\epsilon_{\rm 2-PHAV} = 0.061 \pm 0.006$, $M_{\rm T}=2.92$). For the sake of clarity, in the left panel of Fig.~\ref{nonGunbal} we show the behavior of $\epsilon$ as a function of the mean total energy of the 2-PHAV.  To achieve a more complete characterization of this class of states, we also studied the behavior of the nonG amount by varying the balancing between the two components of the 2-PHAV and keeping the mean number of detected photons of the overall state fixed. The lower bound $\epsilon$ of the nonG amount as a function of the ratio between the two components is plotted in the right panel of Fig.~\ref{nonGunbal}: as one may expect,  it monotonically decreases at increasing the balancing. In fact, the most unbalanced condition reduces to the case in which there is only a single PHAV, whereas the most balanced one corresponds to have a balanced 2-PHAV.
In Fig.~\ref{nonGunbal} we note that the experimental values of $\epsilon$ are larger than those expected from theory and the difference increases at high mean numbers (left panel) or at high unbalancing (right panel).
We ascribe this behavior to small errors (few percent) in the determination of the mean values of detected photons that can occur at relatively high intensities due to some saturation of the detector. Indeed, the sensitivity of $\epsilon$ to the mean number of photons is testified by the error bars associated to the experimental data.
This critical character suggests the exploitation of such nonG measure to test decoy states based on 2-PHAVs \cite{curty09}.

\section{Conclusion}\label{s:remarks}
In conclusion, we have presented an experimental investigation of the non-Gaussian nature of the class of PHAVs by reconstructing the Wigner function of the states in several different conditions. We have also quantified
the nonG amount by exploiting a nonG measure defined in terms of von Neumann entropy and proved the consistency of the two approaches. It is worth noting that the characterization and manipulation of the nonG could be a useful tool for the enhancement of the QKD protocols in which PHAVs are employed.
\par
Moreover, the very good agreement between experimental data and theory reinforces the possibility of using this class of states for further investigations on the nonG of optical states and in particular for the
realization of more complex experimental schemes, also involving nonclassicality. Indeed, the good quality of the results makes our system
particularly useful for quantum state engineering and quantum information processing, despite the many criticalities and difficulties in obtaining a good randomization of the phases, in improving the superposition of more than two light fields (such as in the case of the Wigner function reconstruction of the 2-PHAV) and in maintaining the alignment for hours and for different kinds of measurements.
\par
Finally, it is worth noting that the measurements were actually performed in the mesoscopic photon-number domain, as the reconstruction of the Wigner functions of the states under investigation was achieved by displacing them with a coherent field whose intensity was changed from zero up to four times the mean value of the states themselves.

\section*{Acknowledgements}
The authors thank M.~G.~Genoni for useful discussions.
This work has been supported by MIUR (FIRB ``LiCHIS'' - RBFR10YQ3H).


\begin{thebibliography}{99}
\bibitem{PRL:05} H.-K. Lo, X. Ma, and K. Chen, ``Decoy state quantum key distribution,'' Phys. Rev. Lett. {\bf 94}, 230504 (2005).
\bibitem{APL:07} Y. Zhao, B. Qi, and H.-K. Lo, ``Experimental quantum key
distribution with active phase randomization,'' Appl. Phys. Lett. {\bf
90}, 044106 (2007).
\bibitem{EPJD} H. Inamori, N. L$\ddot{\rm u}$tkenhaus, and D. Mayers, ``Unconditional security of practical quantum key distribution,'' Eur. Phys. J. D {\bf 41,} 599--627 (2007).
\bibitem{CALT:05} H.-K. Lo, and J. Preskill, ``Phase randomization
improves the security of quantum key distribution,'' CALT-68-2556
(2005), arXiv:quant-ph/0504209v1.
\bibitem{curty10} M. Curty, X. Ma, B. Qi, and T. Moroder, ``Passive decoy-state quantum key distribution with practical light sources,'' Phys. Rev. A {\bf 81,} 022310 (2010).
\bibitem{oli:12} S. Olivares, ``Quantum optics in the phase space,''
Eur. Phys. J. Special Topics {\bf 203}, 3--24 (2012).
\bibitem{curty09} M. Curty, T. Moroder, X. Ma, and N. L$\ddot{\rm u}$tkenhaus, ``Non-Poissonian statistics from Poissonian light sources with application to passive decoy state quantum key distribution,'' Opt. Lett. {\bf 34,} 3238--3240 (2009).
\bibitem{lvovsky09} A. I. Lvovsky and M. G. Raymer, ``Continuous-variable optical quantum-state tomography,'' Rev. Mod. Phys. {\bf 81,} 299--332 (2009).
\bibitem{cahill69} K. E. Cahill and R. J. Glauber, ``Density Operators and Quasiprobability Distributions,'' Phys. Rev. {\bf 177,} 1882--1902 (1969).
\bibitem{wallentowitz96} S. Wallentowitz and W. Vogel, ``Unbalanced homodyning for quantum state measurements,'' \pra {\bf 53,} 4528--4533 (1996).
\bibitem{banaszek96} K. Banaszek and K. W$\acute{\rm o}$dkiewicz, ``Direct Probing of Quantum Phase Space by Photon Counting,'' \prl {\bf 76,} 4344--4347 (1996).
\bibitem{zambra07} G. Zambra, A. Allevi, M. Bondani, A. Andreoni, and M. G. A. Paris, ``Nontrivial photon statistics with low resolution-threshold photon counters,'' Int. J. Quantum Inf. {\bf 5,} 305--309 (2007).
\bibitem{GenoniPRA} M. G. Genoni and M. G. A. Paris, ``Quantifying non-Gaussianity for quantum information,'' \pra {\bf 82,} 052341 (2010).
\bibitem{genoni08} M. G. Genoni, M. G. A. Paris, and K. Banaszek, ``Quantifying the non-Gaussian character of a quantum state by quantum relative entropy,'' \pra {\bf 78,} 060303(R) (2008).
\bibitem{OLwigner} M. Bondani, A. Allevi, and A. Andreoni, ``Wigner function of pulsed fields by direct detection,'' \ol {\bf 34,} 1444--1446 (2009).
\bibitem{JMOself} M. Bondani, A. Allevi, A. Agliati, and A. Andreoni, ``Self-consistent characterization of light statistics,'' J. Mod. Opt. {\bf 56,} 226--231 (2009).
\bibitem{andreoni09} A. Andreoni and M. Bondani, ``Photon statistics in the macroscopic realm measured without photon counters,'' \pra {\bf 80,} 013819 (2009).
\bibitem{allevi10} A. Allevi, A. Andreoni, M. Bondani, M. G. Genoni, and S. Olivares, ``Reliable source of conditional states from single-mode pulsed thermal fields
by multiple-photon subtraction,'' \pra {\bf82,} 013816 (2010).
\end{thebibliography}
\end{document}